\documentstyle[12pt,epsf]{article}

\begin{document}

\begin{flushright}
SLAC--PUB--8327\\
February 2000  
\end{flushright}

\begin{center}
{\bf\large
BIT-STRING PHYSICS PREDICTION OF 
$\eta$, THE DARK MATTER/BARYON
RATIO AND $\Omega_M$
\footnote{\baselineskip=12pt
Work supported by Department of Energy contract DE--AC03--76SF00515.}}

\bigskip

H. Pierre Noyes\\
Stanford Linear Accelerator Center\\
Stanford University, Stanford, CA 94309\\
\end{center}

\bigskip\bigskip

\begin{abstract}
Using a simple combinatorial algorithm for generating finite and discrete events 
as our numerical cosmology, we predict that the baryon/photon ratio at the time
of nucleogenesis is $\eta= 1/256^4$, $\Omega_{DM}/\Omega_B= 12.7$ and (for a 
cosmological constant of $\Omega_{\Lambda}=0.6\pm 0.1$ predicted on general grounds by
E.D.Jones) that $0.325 > \Omega_M > 0.183 $. The limits are set not by our theory
but by the empirical bounds on the renormalized Hubble constant of $0.6 < h_0 < 0.8$. 
If we impose the additional empirical bound of $t_0 < 14 \ Gyr$, the predicted
upper bound on $\Omega_M$ falls to $0.26$. The predictions of $\Omega_M$ and $\Omega_{\Lambda}$were in excellent agreement with Glanz' analysis in 1998, 
and are still in excellent agreement with Lineweaver's recent analysis despite the
reduction of observational uncertainty by close to an order of magnitude. 
\end{abstract}

\vfill

\begin{center}
Contributed paper presented at DM2000\\
Marina del Rey, California, February 23-25, 2000.\\
First Afternoon Session, Thursday, February 24 
\end{center}

\newpage

The theory on which I base my predictions is unconventional. Hence it is easier for me 
to show you first the consequences of the predictions in comparison with observation,
in order to establish a presumption that the theory {\it might} be interesting, and then
show you how these predictions came about. 

The predictions are that (a) the ratio of baryons to photons 
was $\eta= 1/256^4 =2.328...\times 10^{-10}=10^{-10}\eta_{10}$
at the time of nucleogenesis, 
(b) $\Omega_{DM}/\Omega_B=127/10=12.7$ and (c) $\Omega_{\Lambda}=0.6$. Comparison 
of prediction (a) with observation is straightforward, as is illustrated in Figure 1.

\begin{figure}[htb]
\begin{center}
\leavevmode
\epsfbox{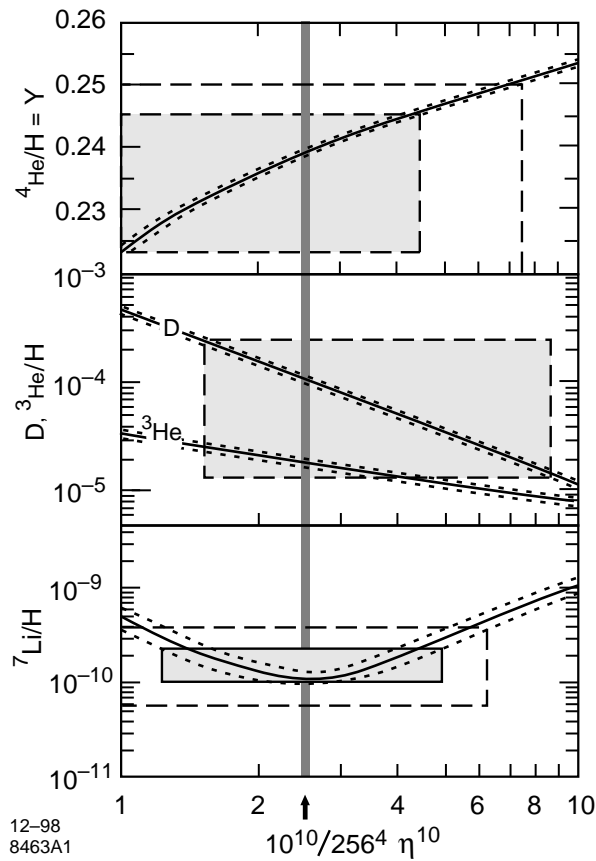}
\end{center}
\caption[*]{Comparison of the bit-string physics prediction that $\eta=
256^{-4}$ with accepted limits on the cosmic abundances as given by
Olive and Schramm in  \cite{PDG98}, p.~119.}
\label{fig1}
\end{figure}

Comparison with observation of prediction (b) that the ratio of dark to baryonic matter is {\it not} straightforward, as was clear at DM98; I suspect that is matter will remain unresolved at this conference (DM2000). However, according to the standard cosmological model, the baryon-photon ratio remains fixed {\it after} nucleogenesis.
In the theory I am relying on, the same is true of the of the dark matter to baryon ratio. Consequently, {\it if} we know the Hubble constant,
{\it and} assume that only dark and baryonic matter contribute, the normalized matter parameter $\Omega_M$ can {\it also} be predicted, as we now demonstrate. 

We know from the currently observed photon density (calculated from the observed $2.728 \ ^oK$ cosmic background radiation) 
that the normalized baryon density is given by \cite{Olive98}
\begin{equation}
\Omega_B = 3.67\times 10^{-3}\eta_{10}h_0^{-2}
\end{equation}
and hence, from our prediction and assumptions about dark matter, that the total mass density will be 13.7 times as large. Therefore we have that
\begin{equation}
\Omega_M= 0.11706 h_0^{-2} \ .
\end{equation}   
Hence, for $0.8 \geq h_0 \geq 0.6$  \cite{Hogan98}, $\Omega_M$ runs from
$0.18291$ to $0.32517$. This clearly puts no restriction
on $\Omega_{\Lambda}$. 

Our second constraint comes from integrating the scaled
Friedman-Robertson-Walker (FRW) equations from a time 
after the expansion becomes matter dominated with no pressure
to the present. Here we assume that this initial time is 
close enough to zero on the time scale of the integration
so that the lower limit of integration can be approximated 
by zero \cite{Primackinp}.
Then the age of the universe as a function of the current values of
$\Omega_M$ and $\Omega_{\Lambda}$ is given by
\begin{eqnarray}
t_0&=&9.77813 h_0^{-1}f(\Omega_M,\Omega_{\Lambda}) \ Gyr\nonumber\\
&=&9.77813 h_0^{-1}f(0.11706h_0^{-2},\Omega_{\Lambda}) \ Gyr
\end{eqnarray} 
where
\begin{equation} 
f(\Omega_M,\Omega_{\Lambda})=\int_0^1dx 
\sqrt{{x\over \Omega_M +(1-\Omega_M-\Omega_{\Lambda})x 
+\Omega_{\Lambda}x^3}} \ .
\end{equation}
For the two limiting values of $h_0$, we see that
\begin{eqnarray}
h_0&=&0.8,\ \ t_0=12.223f(0.18291,\Omega_{\Lambda}) \ \ Gyr \nonumber\\  
h_0&=&0.6,\ \ t_0=16.297f(0.32517,\Omega_{\Lambda}) \ \ Gyr \ .
\end{eqnarray}

\newpage
\begin{figure}[htbp]
\begin{center}
\leavevmode
\epsfbox{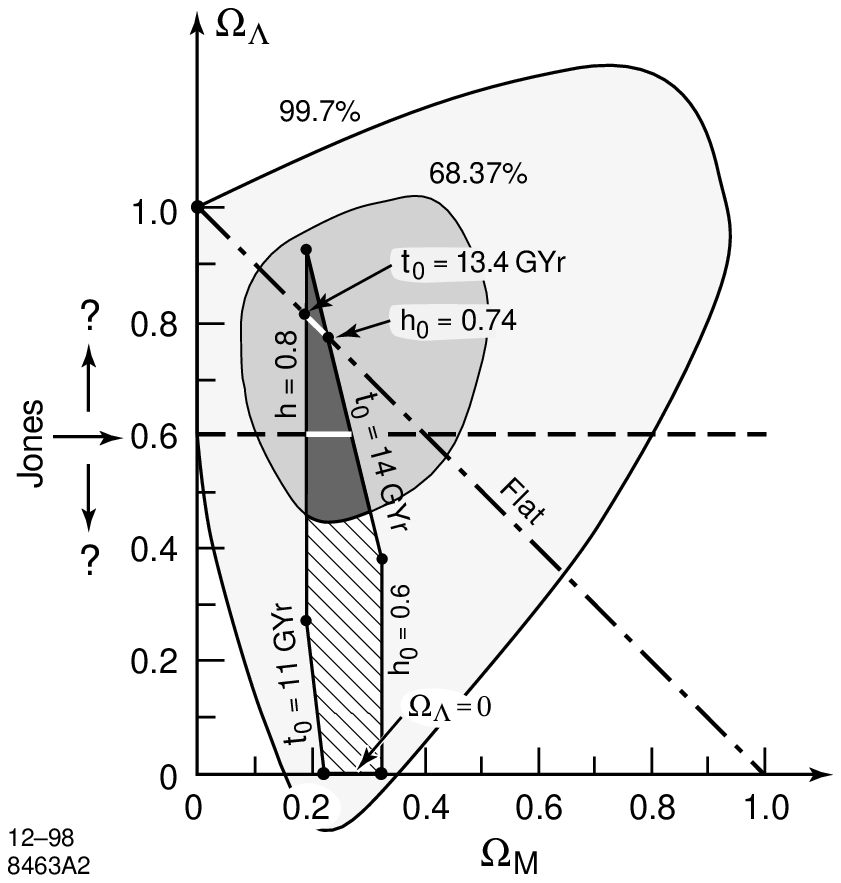}
\end{center}
\caption[*]{Limits on $(\Omega_M, \Omega_{\Lambda})$  set by combining
the Supernovae Type Ia data from Perlmutter, et al. with the Cosmic
Ray Background Experiment (COBE) satellite data as quoted by Glanz
 \cite{Glanz98a} (dotted curves at the 68.37\% and 99.7\% confidence
levels) compared with the predictions of bit-string physics that $\eta_{10} = 10^{10}/256^4$ 
(cf. Fig. 1) and $\Omega_{\rm Dark}/\Omega_B =12.7$. We accept the constraints on 
the scaled Hubble constant $h_0=0.7\pm 0.1$  \cite{Groometal98} and 
on the age of the universe $t_0 = 12.5\pm1.5 \ Gyr$ (solid lines).
We include the predicted constraint $\Omega_{\Lambda} > 0)$. The Jones estimate
of $\Omega_{\Lambda}= 0.6$ is indicated, but the uncertainty was not available in 1998.}
\label{fig2}
\end{figure}

The results are plotted in Figure 2. We emphasize that these predictions were made and published over a decade ago when the observational data were vague and the theoretical climate of opinion was very different from what it is now. The figure just given was presented at ANPA20 (Sept. 3-8, 1998, Cambridge, England) and given wider circulation in\cite{Noyes99}. The calculation (c) that $\Omega_{\Lambda}=0.6$ was
made by Jones before there was any observational evidence for a cosmological constant, let alone a positive one\cite{Jones97}. The precision of the relevant observational limits 
has improved  

\newpage

\begin{figure}[htbp]
\begin{center}
\leavevmode
\epsfbox{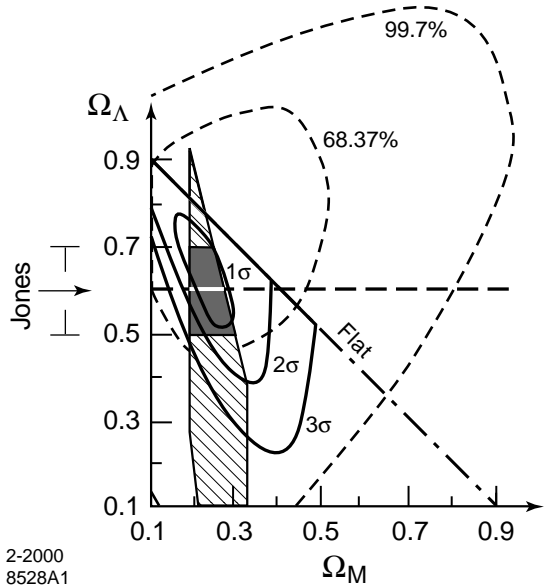}
\end{center}
\caption[*]{Limits on $(\Omega_M, \Omega_{\Lambda})$  set by combining
the Supernovae Type Ia data from Perlmutter, et al. with the Cosmic
Ray Background Experiment (COBE) satellite data as quoted by Glanz
\cite{Glanz98a} (dotted curves at the 68.37\% and 99.7\% confidence
levels) and more recent much improved limits according to Lineweaver\cite{Lineweaver99} (solid curves at $1\sigma, 2\sigma, 3\sigma$) compared with the predictions of bit-string physics that $\eta_{10} = 10^{10}/256^4$ 
(cf. Fig. 1) and $\Omega_{DM}/\Omega_B =12.7$. We accept the constraints on the scaled Hubble constant $h_0=0.7\pm 0.1$  \cite{Groometal98} and 
on the age of the universe $t_0 = 12.5\pm1.5 \ Gyr$ (solid lines).
The Jones estimate of $\Omega_{\Lambda}= 0.6\pm 0.1$ is included.}
\label{fig3}
\end{figure}

\noindent
considerably since DM98. A recent analysis of this new data suitable for our purposes has been made by Lineweaver\cite{Lineweaver99}. His one, two and three $\sigma$ contours are plotted in comparison with the previous observational limits and our (unchanged) earlier predictions in Figure 3. Note how dramatically the regions of uncertainty have shrunken in two years. It is gratifying that our {\it prior} predictions are still close to the center of the allowed region, indicating that it will take a lot more work to show that they are wrong!

The theory I am using has a long history\cite{Noyes97a}, starting with the discovery of the {\it combinatorial hierarchy} in 1961 \cite{Parker-Rhodes62} and the first publication of the work on this idea by Amson, Bastin, Kilmister and Parker-Rhodes in 1966\cite{Bastin66}. The theory is unusual in that it starts from minimal assumptions about what is needed for a physical theory and tries to let the structure of the theory grow out of them.
My own preferred choice of basic assumptions are that a physicist must
(a) be able to tell something from nothing, (b) be able to tell whether things are the same or different, and (c) must assume a basic arbitrariness in the universe which underlies the stochastic effects exhibited by quantum events. I further assume that we should use the simplest possible mathematical structures to model and develop these concepts. (a) is simply modeled by bit multiplication; (b) is simply
modeled by bit addition (addition modulo 2, XOR, symmetric difference,...) or, as it is referred to in the ANPA program, {\it discrimination}.

The third requirement, together with the usual scientific assumption that we can keep {\it historical records} and examine them at later times,
is accomplished by constructing a computer model called
{\it program universe}\cite{Manthey86,Noyes&McGoveran89,Noyes97b}
which yields a growing universe of ordered strings of the {integers} ``0'' and ``1''.  Here we remind the reader of how we use {\it discrimination} (``$\oplus$'') between ordered strings of zeros and ones ({\it bit-strings}) defined by
\begin{equation}
({\bf a}(W)\oplus {\bf b}(W))_w =(a_w-b_w)^2;\ \
a_w,b_w \in 0,1;\ \ w \in 1,2,....,W
\end{equation}   
to generate a growing universe of bit-strings which at each step
contains $P(S)$ strings of length $S$. 
The algorithm is very simple, as can be seen from the flow diagram in Fig. 4. We start with a rectangular block of rows and columns containing 
only the bits ``0'' and ``1''. We then pick two rows arbitrarily and if their discriminant is non-null, adjoin it to the table as a new row.
If it is null, we simply adjoin an arbitrary column (Bernoulli sequence) 
to the table and recurse to picking two arbitrary rows. That this model 
contains arbitrary elements and (if interpretable in terms of known aspects of the practice of physics) an historical record (ordered by the number of TICK's, or equivalently by the row length) should
be clear from the outset. The forging of rules that will indeed connect the model to the {\it actual} practice of physics is the primary problem that has engaged me ever since the model was created.  

\begin{figure}[htbp]
\begin{center}
\leavevmode
\epsfbox{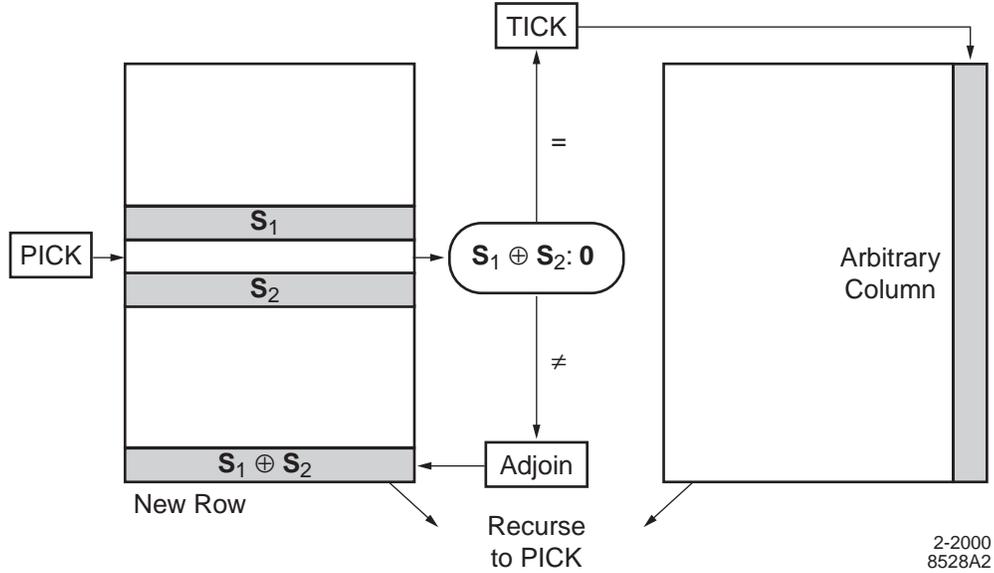}
\end{center}
\caption[*]{Flow diagram for constructing a bit-string universe growing by one row or one column at a time (see text).}
\label{fig4}
\end{figure}

Program universe provides
a separation into a conserved set of ``labels'', and a growing set of
``contents'' which can be thought of as the space-time ``addresses''
to which these labels refer. To see this, think of all the left-hand,
finite length $S$ portions of the strings which exist when the program
TICKs and the string-length goes from $S$ to $S+1$. Call these {\it
labels} of length $L=S$, and the number of them at the critical 
tick $N_0(L)$. Further PICKs and
TICKs can only add to this set of labels those which can be produced
from it by pairwise discrimination, with no impact from the (growing in length
and number) set of content labels with length  $S_C=S-L > 0$. 
If $N_I \leq N_0(S_L)$ of these labels are {\it discriminately
independent}, then the maximum number of distinct labels they
can generate, no matter how long program universe runs,
will be $2^{N_I} - 1$, because this is the maximum number of ways we can
choose combinations of $N_I$ distinct things taking them $1,2,...,N_I$  
times. We will interpret this fixed number of possibilities as a 
representation of the  quantum numbers of systems of
 ``elementary particles'' allowed in our bit-string universe 
and use the growing content-strings to represent 
their (finite and discrete) locations
in an expanding space-time description of the universe.

This label-content schema then allows us to interpret the events
which lead to  TICK as four-leg Feynman diagrams representing
a stationary state scattering process. Note that for us to
find out that the two strings found by PICK are the same,
we must either pick the same string twice or at some previous step 
have produced (by discrimination) and adjoined the string
which is now the same as the second one picked. Although
it is not discussed in bit-string language, a little thought
about the solution of a relativistic three body scattering
problem Ed Jones and I have found \cite{Noyes&Jones2000}
shows that the driving term ($>-<\atop -$) is always a four-leg Feynman
diagram ($>-<$) plus a spectator ($\ - \ $) whose quantum numbers are 
{\it identical} with the
quantum numbers of the particle in the intermediate state 
connecting the two vertices. The 
step we do not take here is to show that the labels do
indeed represent quantum number conservation and the contents
a finite and discrete version of relativistic 
energy-momentum conservation. But we hope that enough has been said
to show how we could interpret program universe as representing
a sequence of contemporaneous scattering processes, and an algorithm
which tells us how the space in which they occur expands.  

Short-circuiting and reordering the actual route by which 
my current interpretation of this model was arrived at, we note that the two
basic operations in the model which provide locally novel bit-strings (Adjoin
and TICK) are isomorphic, respectively, to a three-leg or a four-leg Feynman diagram. This is illustrated in Fig. 5. Note that the internal (exchanged particle) state in the Feynman
diagram is {\it necessarily} accompanied by an identical (but distinct) ``spectator'' somewhere else in the (coherent) memory.
\begin{figure}[htbp]
\begin{center}
\leavevmode
\epsfbox{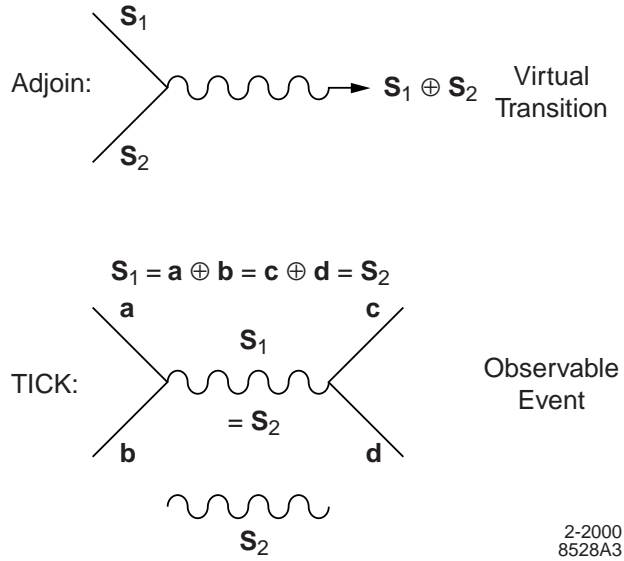}
\end{center}
\caption[*]{Interpretation of the Adjoin and TICK operations of Program Universe as Feynman diagrams(see text).}
\label{fig5}
\end{figure}

We do not have space here to explain how, in the more detailed dynamical
interpretation, the three-leg diagrams conserve (relativistic) 3-momentum
but not necessarily energy  (like vacuum fluctuations) while the four-leg diagrams conserve both
3-momentum and energy and hence are candidates for potentially observable events. We are particularly pleased that the observable events created
by Program Universe necessarily provide {\it two} locally identical but distinct strings (states) because these are the starting point for a relativistic finite particle number quantum scattering theory which has
non-trivial solutions\cite{Noyes&Jones2000}. But we {\it do} need to explain how this interpretation of program universe does connect up with the work on the combinatorial hierarchy. 

At this point we need a guiding principle to show us how we can
``chunk'' the growing information content provided by the discriminate
closure of the label portion of the strings in such a way as to generate a hierarchical representation of the quantum numbers that these labels represent. Following a suggestion of David McGoveran's \cite{McGoveran98},
we note that {\it we can guarantee that the representation has a
coordinate basis and supports linear operators
by mapping it to square matrices}. 

The mapping scheme originally used
by Amson, Bastin, Kilmister and Parker-Rhodes \cite{Bastin66}
satisfies this requirement. This scheme requires us to
introduce the multiplication operation ($0\cdot 0=0=0\cdot 1=0=1\cdot
0$, $1\cdot 1=1$), converting our bit-string formalism into the {\it
field} $Z_2$. First note, as mentioned above, that any set of $n$
discriminately independent ({\it d.i.}) strings will generate exactly $2^n -1$
discriminately closed subsets ({\it dcss}). Start with two d.i. strings ${\bf a}$,
${\bf b}$. These generate three d.i. subsets, namely        
$\{ {\bf a} \}$, $\{ {\bf b} \}$, 
$\{ {\bf a},{\bf b},{\bf a}\oplus {\bf b} \}$. Require each dcss 
(\{ $\ $ \}) to contain only the eigenvector(s),
of three $2\times 2$ {\it mapping matrices} which 
(1) are non-singular (do not map onto zero) and (2) are d.i.
Rearrange these as strings. They will then generate seven dcss.
Map these by seven d.i. $4\times 4$ matrices, which meet the same
criteria (1) and (2) just given. Rearrange these as  seven d.i. strings
of length 16. These generate $127=2^7-1$ dcss. These can be mapped
by 127 $16\times 16$ d.i. mapping matrices, which, rearranged
as strings of length 256, generate
$2^{127}-1 \approx 1.7\times 10^{38}$ dcss. But these cannot be
mapped by $256\times 256$ d.i. matrices because there are at most
$256^2$ such matrices and $256^2 \ll 2^{127}-1$. 
Thus this {\it combinatorial hierarchy} terminates at the fourth
level. The mapping matrices are not unique, but exist, as has been
proved by direct construction and an abstract proof \cite{Bastinetal79}.
It is easy to see that the four level hierarchy constructed by these
rules is {\it unique} because starting with d.i. strings of length 3 or
4 generates only two levels and the dcss generated by d.i. strings of
length 5 or greater cannot be mapped. 

Making physical sense out of these numbers is a long story \cite{Noyes97a},
and making the case that they give us the quantum numbers of the
standard model of quarks and leptons with exactly 3 generations
has only been sketched \cite{Noyes94}. However we do not require
the completely worked out scheme to make interesting cosmological
predictions.  The ratio of dark to ``visible''  (i.e. electromagnetically
interacting) matter is the easiest to see. The electromagnetic
interaction first comes in when we have constructed the first three
levels giving 3+7+127 =137 dcss, one of which is identified with
electromagnetic interactions because it occurs with probability 
$1/137 \approx e^2/\hbar c$. But the construction must first complete
the first two levels giving 3+7=10 dcss. Since the construction is
``random'' and this will happen
many, many times as program universe grinds along, we will
get the 10 non-electromagnetically interacting labels 127/10 times
as often as we get the electromagnetically interacting labels.
Our prediction of $M_{DM}/M_{B}= 12.7$ is that naive. 
 
The $1/256^4$ prediction for $N_{B}/N_{\gamma}$ is comparably naive.
Our partially worked out scheme of relating bit-string events
to particle physics \cite{Noyes94,Noyes97a}, makes it clear that
photons, both as labels (which communicate with particle-antiparticle
pairs) and as content strings will contain equal numbers of zeros
and ones in appropriately specified portions of the strings.
Consequently they can be readily identified as the most probable 
entities in any assemblage of strings generated by whatever pseudo-random
\newpage

\begin{figure}[htb]
\begin{center}
\leavevmode
\epsfbox{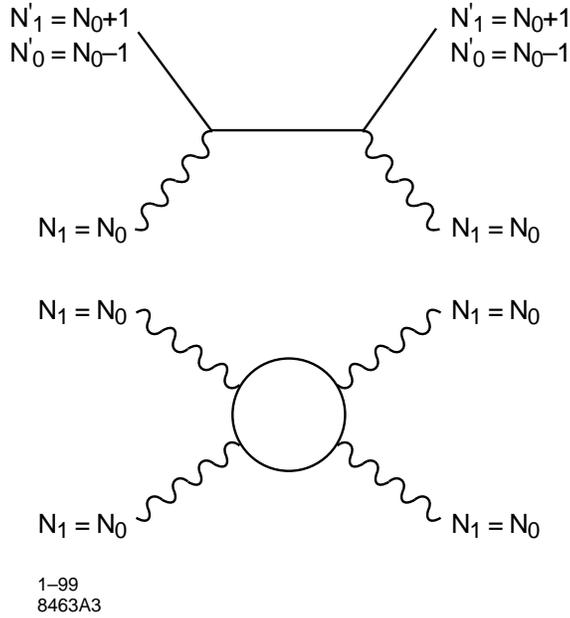}
\end{center}
\caption[*]{Comparison of bit-string labeled processes after the label
length is fixed at 256 interpreted as baryon ($N_1'=N_0+1$) 
photon ($N_1=N_0$) and photon-photon scattering. Here $N_1$ 
and $N_0$ symbolize, respectively,  the number of ones and zeros in the 
label part of the string  (which is of length 256). Program universe guarantees 
that, in the absence of further considerations, the content part of the strings 
will have an equal number of zeros and ones with very high probability as 
the string length (universe) grows. }
\label{fig6}
\end{figure}

\noindent
number generator is used to construct the arbitrary actions and bit-strings needed in actually running program universe.
This scheme also makes the simplest representation of fermions and
anti-fermions contain one more ``1'' and one less ``0'' than the photons (or {\it visa versa}).
(Which we call ``fermions'' and which ``anti-fermions'' is, to begin
with, an arbitrary choice of nomenclature.) Since our dynamics insures
conventional quantum number conservation by construction, the problem
--- as in conventional theories---is to show how program universe
introduces a bias between ``0'' 's and ``1'' 's
once the full interaction scheme is developed. 

Since program universe has to start out with two
strings, and both of these cannot be null if the evolution is lead
anywhere, the first significant PICK and discrimination will
necessarily lead to a universe with three strings, two of which are
``1'' and one of which is ``0''. Subsequent PICKs and TICKs are
sufficiently ``random'' to insure that (at least statistically)
there will be an equal number of zeros and ones, apart from the initial
bias giving an extra one. Once the label length of 256 is reached,
and sufficient space-time structure (``content strings'') generated and
interacted to achieve thermal equilibrium, this label bias for a 1
compared to equal numbers of zeros and ones will persist for 1 in
256 labels. But to count the equilibrium processes relevant 
to computing the ratio of baryons to photons, we must compare the
labels leading to baryon-photon scattering compared to those
leading to photon-photon scattering. This requires the
baryon bias of 1 to appear in one and only one of the four initial
(or final, since the diagrams are time symmetric) state labels
of length 256 involved in that comparison; the two relevant diagrams are 
illustrated in Fig.6 \ref{fig1}, which assumes that the above mentioned
interpretation of the strings causing observable TICK's as
four  leg Feynman diagrams has been satisfactorily demonstrated. 
As a trivial example take the baryon-antibaryon-photon vertex to be
${\bf B}\oplus {\bf \bar{B}} \oplus {\bf \gamma} = {\bf 0}$ with
${\bf B}=(1110)$, ${\bf \bar{B}}=(0010)$ and ${\bf \gamma}= (1100)$. We conclude that, in the absence of further information, 
$1/256^4$ is the program universe prediction
for the baryon-photon ratio at the time of big bang nucleosynthesis.

Since Jones' paper \cite{Jones97} is still in preparation, I am at liberty here only to quote:
\begin{quote}
From general operational arguments, Ed Jones has shown how to start
from $\sim N$ Plancktons and self-generate a universe with $\sim N'$
baryons which---for appropriate choice of $N$---resembles our
currently observed universe. In particular it must necessarily have a 
positive cosmological constant characterized by $\Omega_{\Lambda} \sim
0.6\pm 0.1$. 
\end{quote} 

We note further that Jones' general arguments a) are completely
compatible with {\it program universe} and b) do not in themselves
fix the value of $N$. Further, the estimate given above, which
was made
before and independent of the calculations reported in the last
section, fell squarely in the middle of the region allowed in 1998
(see Fig. 2), and continues to do so despite the remarkable progress that has been made since DM98 (see Fig.3). Clearly, pursuing the combination of these two lines of reasoning could prove to be very exciting.

\begin{center} 
{\bf APPENDIX}
\end{center}

In order to underpin our claim that we can model a finite particle number version of relativistic quantum mechanics with particle creation, etc.
using bit-strings we give on the next page the predictions of coupling constants and mass ratios calculated using our theory. As in any mass, length, time theory we are allowed three empirical, dimensional constants
which are measured by standard techniques to connect our abstract theory
to measurement. These we take to be the mass of the proton $m_p$, Planck's constant $\hbar$ and the velocity of light $c$. Everything else is calculated. Agreement with observation, given on the next page, is not perfect; we believe it is impressive. For more detail see\cite{Noyes97a}.

A tentative bit-string representation of the quantum numbers of the (three generation) standard model of quarks and leptons is given on the following page (Fig. 7).
\newpage

$$G_N^{-1} {\hbar c \over m_p^2}=
[2^{127} + 136]\times
{\bf [1-{1\over 3\cdot 7\cdot 10}]} =1.693 \ 31\ldots\times 10^{38}$$
$$ experiment =1.693 \ {\bf 58}(21) \times 10^{38}$$
$$\alpha ^{-1} (m_e)=
137\times{\bf [1- {1 \over 30 \times 127}]^{-1}} =
137.0359 \ {\bf 674....} $$
$$  experiment =137.0359\ 895(61)$$
$$G_Fm_p^2/\hbar c
= [256^2\sqrt{2}]^{-1}\times {\bf [1 - {1 \over 3\cdot 7}]}
=1.02 \ {\bf 758\ldots}\times 10^{-5}$$
$$ experiment = 1.02 \ 682(2)\times 10^{-5}$$
$$sin^2\theta _{Weak}=
0.25{\bf [1 - {1 \over 3\cdot 7}]^2}
= 0.2267\ldots$$
$$experiment =0.22{\bf 59}(46)]$$
\begin{equation}
{m_p\over m_e} ={137 \pi\over <x(1-x)><{1\over y}>}=
{137 \pi\over ({3\over 14})[1 + {2\over 7} + {4\over 49}]({4\over 5})}
 = 1836.15 \ {\bf 1497\ldots} 
\end{equation} 
$$ experiment =1836.15\ 2701(37)$$
$$m_{\pi}^{\pm}/ m_e
=275{\bf [1 - {2\over 2\cdot 3 \cdot 7\cdot 7}]}
= 273.12 \ {\bf 92\ldots}$$
$$ experiment = 273.12 \ 67(4)$$
$$m_{\pi ^0}/m_e
=274 {\bf [1- {3\over 2\cdot 3 \cdot 7 \cdot 2}]}
=264.2 \ {\bf 143\ldots}$$
$$experiment=264.1 \ 373(6)]$$
$$m_{\mu}/m_e
=3\cdot7\cdot10[1-{3\over 3\cdot7\cdot10}]= 207$$
$$ experiment = 206.768 \ 26(13)$$
$$G^2_{\pi N\bar N}=
[({2 M_N\over m_{\pi}})^2-1]^{{1\over 2}}=[195]^{{1\over 2}}=13.96....$$
$$experiment = 13.3(3), \ or \ greater \ than \ 13.9 $$

\newpage
\begin{figure}[htb]
\begin{center}
\leavevmode
\epsfbox{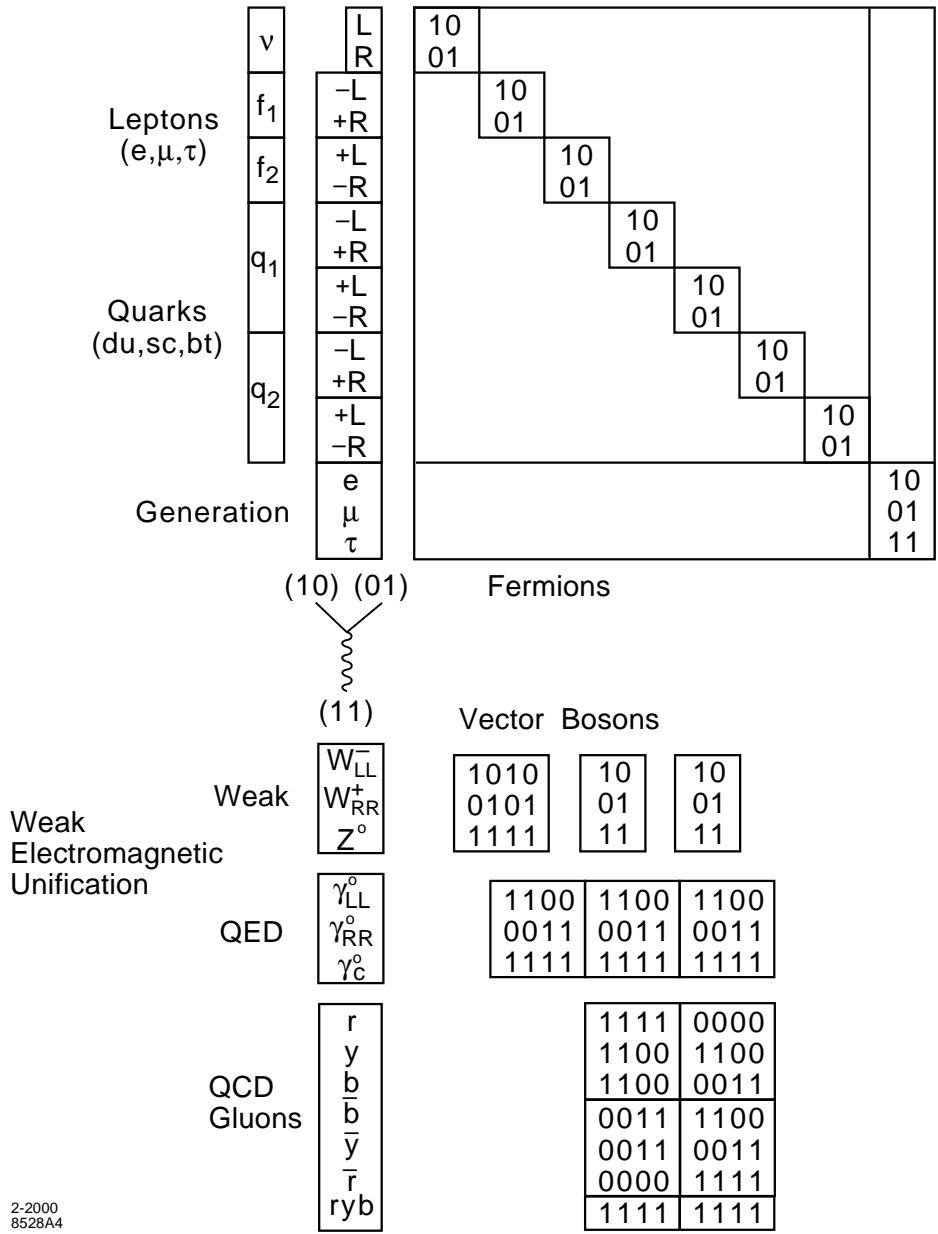}
\end{center}
\caption[*]{Skeleton of a label scheme for labels of length 16 which conveys the same quantum number information as the standard model of quarks and leptons.}
\label{fig7}
\end{figure}

\end{document}